\documentclass[aps,prc,twocolumn,10pt, superscriptaddress,floatfix]{revtex4-1}
\usepackage{amssymb,epsfig}
\usepackage{amsmath,bm}
\usepackage{mathtools}
\newcommand{\be}{\begin{equation}}
\newcommand{\ee}{\end{equation}}
\newcommand{\bea}{\begin{eqnarray}}
\newcommand{\eea}{\end{eqnarray}}
\newcommand{\mathscr}{\cal}

\begin{document}

\title{Impact of complex many-body correlations on electron capture in thermally excited nuclei around $^{78}$Ni}
\author{Elena Litvinova}
\affiliation{Department of Physics, Western Michigan University, Kalamazoo, MI 49008, USA}
\affiliation{National Superconducting Cyclotron Laboratory, Michigan State University, East Lansing, MI 48824, USA}
\affiliation{GANIL, CEA/DRF-CNRS/IN2P3, F-14076 Caen, France}
\author{Caroline Robin}
\affiliation{Fakult\"at f\"ur Physik, Universit\"at Bielefeld, D-33615 Bielefeld, Germany}

\date{\today}

\begin{abstract}
We link complex many-body correlations, which play a decisive role in the structural properties of atomic nuclei, to the electron capture occurring during star evolution.
The recently developed finite-temperature response theory, taking into account the coupling between single-nucleon and collective degrees of freedom, is applied to spin-isospin transitions, which dominate the electron capture rates. Calculations are performed for $^{78}$Ni and for the surrounding even-even nuclei associated with a high-sensitivity region of the nuclear chart in the context of core-collapse supernova simulations. The obtained electron capture rates are compared to those of a simpler thermal quasiparticle random phase approximation (TQRPA), which is standardly used in such computations. The comparison indicates that correlations beyond TQRPA lead to significantly higher electron capture rates under the typical thermodynamical conditions.    

\end{abstract}

\maketitle

\section{Introduction} 
\label{Introduction}

The core collapse of massive stars with the subsequent supernova explosion are among the most spectacular cataclysmic events in nature. 
It has been established that the core evolution is governed essentially by weak interaction processes in nuclei, such as electron capture and beta decay \cite{Bethe1979}, 
and their accurate knowledge is needed for reliable astrophysical modeling. The rates of the corresponding nuclear reactions can be extracted from the nuclear transitions with spin and isospin transfer, among which the transitions of the Gamow-Teller (GT) type are known to play the dominant role. The spin-isospin excitations with additional angular momentum transfer may, however, contribute significantly at certain thermodynamical conditions, see, for instance, Refs. \cite{Niu2011,Dzhioev2019,Ravlic2020,Langanke2020} and references therein for more details. 

As a massive corpus of the weak reaction rates is required for successful simulations of various stages of the star evolution, neither existing nor prospective experimental data is capable of providing a complete input. Moreover, even for the nuclear systems, which are accessible at the existing experimental nuclear physics facilities, the thermodynamical conditions of the stellar environment can not be reproduced in laboratory. Therefore, accurate theoretical predictions of weak interaction processes in a wide range of nuclear masses, isospins and temperatures become extremely important.

However, as the nuclear many-body problem is outstandingly complicated and still far from being solved, only very approximate calculations are available, whose degree of success is difficult to judge because of shortage of experimental data. After the first tabulation of Refs. \cite{Fuller1980,Fuller1982,Fuller1985} obtained by making use of a simple independent particle model, the nuclear shell-model tabulations became available for $pf$-shell nuclei \cite{Dean1998,Langanke2001,Suzuki2011,Mori2016}. In principle, the shell model is capable of providing most accurate calculations for low-energy nuclear excited states, however, a non-universal character of its Hamiltonians as well as its limitations on the excitation energies and nuclear mass calls for alternative methods. The approaches based on the quantum-mechanical equations of motion (EOM) and modern universal density functionals represent another class of models. However, the majority of presently available calculations of this type are limited by the simplest random phase approximation (RPA) or its superfluid version, quasiparticle RPA (QRPA) \cite{Borzov:2003bb,Niu2011,Dzhioev2019,Dzhioev2020,Ravlic2020}. These approximations are free of the limitations inherent in the shell model, however, they are confined by the too simple ansatz of the nuclear wave functions and, therefore, fail in describing spectral characteristics with the quality necessary for the extraction of beta decay and electron capture rates even at zero temperature, as far as this can be tested experimentally. 

The way out of this problem becomes evident if one considers extensions of RPA with dynamical kernels in the equation of motion for the particle-hole correlation function. Indeed, RPA is known to keep only the static part of the kernel and to neglect completely its dynamical part. This fact is responsible for poor performance of RPA in nuclear physics. The dynamical kernel contains complex correlations and can not be treated exactly without referring to the equations of motion for higher-rank correlation functions. Instead, the dynamical kernel itself admits various approximations, among which the most common are known as second RPA (SRPA), particle-vibration coupling (PVC), multiphonon approach etc. \cite{LitvinovaSchuck2019}. While SRPA treats the dynamical kernel in the lowest order in the nucleon-nucleon interaction and, thus, includes the minimal degree of correlations, the PVC and the multiphonon approaches contain  non-perturbative resummations capturing the emergent collective phenomena and, thus, are more adequate for nuclear applications. Indeed, various versions of the PVC approach based on non-relativistic \cite{NiuColoBrennaEtAl2012,Severyukhin:2014kja,NiuNiuColoEtAl2015,NiuColoVigezziEtAl2014,Niu2018} and relativistic \cite{MarketinLitvinovaVretenarEtAl2012,LitvinovaBrownFangEtAl2014,RobinLitvinova2016,RobinLitvinova2018} effective interactions showed a considerable improvement over (Q)RPA in the description of nuclear spin-isospin excitations. 

The latter approaches combine the advantage of being not limited by nuclear mass and excitation energy with the strength of the shell model generating rather complex wave functions of the excited states. However, their leading approximation to PVC is confined by configurations where only one phonon mode is exchanged between nucleons at a time - these configurations are not sufficient to reproduce the observed fine features of nuclear spectra and their level densities. Another simplification is the treatment of the ground state wave function, which is typically assumed to be of the Hartree or Hartree-Fock type. Recent developments of Refs. \cite{LitvinovaSchuck2019,Robin2019} are designed to overcome these limitations.

Since atomic nuclei in stars are embedded in hot environments, thermal population of nuclear excited states should be taken into account in calculations of the reaction and decay rates.
This implies that not only transitions from ground to excited states, but also transitions between excited states contribute to the electron capture and beta decay rates. While the shell model approach can explicitly access those transitions through the global diagonalization procedures, the EOM-based models should be completely redesigned to be able to generate such transitions. 
However, under the condition of sufficiently high level density, such approach as (Q)RPA can be straightforwardly generalized for initial states of the thermal mean-field character \cite{Sommermann1983,RingRobledoEgidoEtAl1984}. This fact is employed, for instance, in calculations of electron capture of Refs. \cite{Niu2011,Dzhioev2010,Dzhioev2019,Dzhioev2020,Ravlic2020,Nabi2019}.
Finite-temperature extensions of the approaches with EOM dynamical kernels are essentially more complicated although some implementations for non-charge-exchange nuclear transitions became available earlier in the framework of the nuclear field theory (NFT) \cite{BrogliaBortignonBracco1992}. Recently, a relativistic version of thermal NFT was formulated and implemented for calculations of beta decay rates in hot environments \cite{LitvinovaRobinWibowo2020}. This first finite-temperature relativistic QFT approach to nuclear spin-isospin excitations is based on an accurate solution of the EOM for the proton-neutron response function with both static and dynamic kernels, where the latter includes the leading PVC contributions, and the former is represented by the pion and rho-meson exchanges. In this approach named finite-temperature proton-neutron relativistic time blocking approximation (FT-pnRTBA), the effects of finite temperature on the relativistic mean field, single-particle states, phonon modes and PVC are taken into account fully self-consistently. Calculations were performed for both allowed and forbidden $\beta^{-}$ Gamow-Teller transitions in the typical r-process waiting-point nuclei, and the relative roles of these transitions in beta decay as functions of temperature were analyzed.

In this work, we investigate the potential of the FT-pnRTBA to describe $\beta^{+}$ Gamow-Teller transitions and the associated electron capture rates in neutron-rich nuclei around $^{78}$Ni, which are abundantly produced in core-collapse supernovae at certain thermodynamical conditions. Moreover, Refs. \cite{Sullivan2016,Titus2018,Raduta2017,Pascal2020} concluded that nuclei around N = 50 represent the high-sensitivity region in the context of simulating and understanding the evolution of core-collapse supernovae. 
A good reproduction of GT strength and beta decay rates for many nuclear systems, including the Ni isotopic chain \cite{RobinLitvinova2016,LitvinovaRobinWibowo2020}, serves as a benchmark for the finite-temperature extension of our approach, which uses only the parameters of the covariant energy density functional adjusted globally to nuclear masses and radii \cite{VretenarAfanasjevLalazissisEtAl2005}. 

Section \ref{Formalism} of this article is focused on  the main building blocks of our formalism, Section \ref{Results} provides details of calculations and presents results and discussion, and Section \ref{Summary} concludes on the advancements made in this work.

\section{Model assumptions and formalism}
\label{Formalism}

The initial assumption for the following method is that the atomic nucleus embedded in a hot environment can be described statistically as a compound nucleus and assigned with thermodynamical characteristics, such as temperature $T$ and entropy $S$ \cite{Sommermann1983}. 
One has to perform minimization of the grand potential $\Omega(\mu,T)$ 
\be
\Omega(\mu,T) = E  - T S - \mu N
\label{Omega}
\ee
with the Lagrange multipliers $\mu$ and $T$,
average energy $E$ and particle number $N$. The 
minimization defines the density operator $\hat{\rho}$ which enters the thermal averages:
\be
S = -k\text{Tr}({\hat{\rho}}\text{ln}\hat{\rho}), \ \ \ 
\ N= \text{Tr}(\hat{\rho} {\hat{\cal N}}),
\label{SN}
\ee 
with $\hat{\cal N}$ being the particle number operator, and $k$ the Boltzmann constant equal to one in the natural units. To describe the energy we employ the covariant functional of the nucleonic density and classical meson and photon fields $\phi_m$ \cite{VretenarAfanasjevLalazissisEtAl2005}:
\bea
E[\hat{\rho},\phi_m] &=& \text{Tr}[({\vec\alpha}\cdot{\vec p} + \beta M)\hat{\rho}] + \sum\limits_m\Bigl\{\text{Tr}[(\beta\Gamma_m\phi_m)\hat{\rho}] \pm \nonumber \\
&\pm& \int d^3r \Bigl[\frac{1}{2} ({\vec\nabla}\phi_m)^2 + U(\phi_m)\Bigr]\Bigr\}.
\label{cedf}
\eea
In Eq. (\ref{cedf})  $M$ is the nucleon mass, and  $U(\phi_m)$ denote the non-linear sigma-meson potentials \cite{Lalazissis1997} . The upper "+" sign  corresponds to the scalar $\sigma$-meson, and the lower "-" sign stands for the vector $\omega$-meson, $\rho$-meson and photon, while the index "$m$" runs over the bosonic and Lorentz indices \cite{VretenarAfanasjevLalazissisEtAl2005}.
 
The eigenvalues of the nucleonic density are given by the Fermi-Dirac distribution:
\be
n_1(T) = n(\varepsilon_1, T) = \frac{1}{1 + \text{exp}\{\varepsilon_1 /T \}},
\label{FermiDirac}
\ee
where the number index runs over the complete set of the single-particle quantum numbers in the Dirac-Hartree basis. 
The single-particle energies $\varepsilon_1 = {\tilde\varepsilon_1} - \mu_{\tau_1}$ are measured from the chemical potential $\mu_{\tau_1}$ of the subsystem with the given isospin $\tau_1$.
%


For transitions without particle transfer, the spectral properties, such as excitation energies and transition amplitudes, are determined by
the particle-hole response function ${\cal R}(14,23)$ which, in general, satisfies the Bethe-Salpeter equation (BSE) \cite{Salpeter1951}:
\bea
{\cal R}(14,23) &=& {\cal G}(1,3){\cal G}(4,2) + \nonumber \\
&+& \sum\limits_{5678} {\cal G}(1,5){\cal G}(6,2)V(58,67){\cal R}(74,83)
\label{bse1}
\eea
transformed according to Matsubara's prescription \cite{Matsubara1955}.
Namely, ${\cal G}(1,3)$ are the Matsubara temperature Green's functions of single particles defined for the imaginary time differences: $t_{13} = t_1-t_3$ ($0<t_{1,3}<1/T$) \cite{Matsubara1955,Abrikosov1965} and the number indices run over the single-particle variables and time: $1 = \{k_1,t_1\}$. 

The interaction kernel $V(58,67)$ of Eq. (\ref{bse1}) contains all the in-medium physics and can be treated in various approximations. Most generally, in the case of the presence of only two-body forces and with the local character of the external fields, it is split into an instantaneous (static) term given by a contraction of the vacuum fermion-fermion interaction with the two-fermion density and a time-dependent (dynamic) term, where the fully correlated two-particle-two-hole ($2p2h$) fermionic propagator is twice contracted with the fermion-fermion interaction matrix elements, see, for instance,  Ref. \cite{LitvinovaSchuck2019} for details. In the approaches based on effective interactions, the static term can be approximated by the in-medium effective interaction and the dynamic term can be either neglected, like in (Q)RPA, or treated approximately with varied degree of accuracy. In this work, we use  the PVC model in the leading approximation, where the correlated $2p2h$ propagator is factorized into an uncorrelated $1p1h$ and a correlated $1p1h$ ones. The correlated $1p1h$ propagator is often called phonon (vibration), so the approximation to the dynamic kernel is encoded in $1p1h\otimes phonon$ configurations.

Under this assumption,  Eq. (\ref{bse1}) can be transformed to an equation replacing the complete Matsubara temperature Green's functions ${\cal G}(1,3)$ by the mean-field ones ${\tilde{\cal G}}(1,3)$,
which are connected via the Dyson equation:
\be
{\cal G}(1,2) = {\tilde{\cal G}}(1,2) + \sum\limits_{1'2'}{\tilde{\cal G}}(1,1')\Sigma^e(1'2'){\cal G}(2',2)
\ee
with the dynamic self-energy $\Sigma^e$ containing the PVC effects in the leading approximation, see Ref.  \cite{WibowoLitvinova2019} for details.
Then, introducing the uncorrelated particle-hole propagator ${\tilde{\cal R}}(14,23) = {\tilde{\cal G}}(1,3){\tilde{\cal G}}(4,2)$
and the redefined interaction kernel ${\cal W}(14,23)$, Eq. (\ref{bse1}) can be transformed to the following form:
\be
{\cal R}(14,23) = \tilde{\cal R}(14,23) + \sum\limits_{5678} \tilde{\cal R}(16,25){\cal W}(58,67){\cal R}(74,83).
\label{bse2}
\ee
Here the uncorrelated particle-hole propagator ${\tilde{\cal R}}(14,23)$  in the time domain is a product of two fermionic temperature mean-field Green's functions $\widetilde{\mathcal{G}}$
which, in the imaginary-time representation, read \cite{Abrikosov1965}:
\begin{eqnarray}
\label{Full Mean-Field}\widetilde{\mathcal{G}}(2,1)&=&\sum_{\sigma}\widetilde{\mathcal{G}}^{\sigma}(2,1),\\
\label{Component Mean-Field}\widetilde{\mathcal{G}}^{\sigma}(2,1)&=&-\sigma\delta_{{1}{2}}n(-\sigma\varepsilon_{{1}},T)
e^{-\varepsilon_{{1}}t_{21}}\theta(\sigma t_{21}),
\end{eqnarray}
where $t_{21}=t_{2}-t_{1}$ ($-1/T<t_{21}<1/T$), $\theta(t)$ is the Heaviside step-function and
the index $\sigma=+1(-1)$ denotes the forward (backward) component of $\widetilde{\mathcal{G}}$. 
The new interaction kernel ${\cal W}$ decomposes as follows:
\bea
{\cal W}(14,23) = {\tilde V}(14,23) + V^e(14,23) +  \nonumber \\ 
+ {\tilde{\cal G}}^{-1}(1,3)\Sigma^e(4,2) +  \Sigma^e(1,3){\tilde{\cal G}}^{-1}(4,2),
\label{W-omega}
\eea
into the static interaction $\tilde V$ specified below, the phonon-exchange term $V^e$ and the corresponding self-energy terms ${\tilde{\cal G}}^{-1}\Sigma^e$ and $\Sigma^e{\tilde{\cal G}}^{-1}$, such that
$V^e =\delta \Sigma^e / \delta{\tilde{\cal G}}$, in analogy to the BSE in the particle-hole channel at $T=0$ \cite{Tselyaev1989,Tselyaev2007,LitvinovaRingTselyaev2007}. In the kernel of Eq. (\ref{W-omega}) we used an additional assumption of a smallness present in the dynamic self-energy, so that the latter is kept only in linear order.

In the majority of applications with sufficiently weak external fields, in order to calculate the observed excitation spectra, the
response function has to be eventually contracted with external field operators of local and instantaneous character. This implies that in the final expressions only two time variables in the response function survive. The time blocking approximation, which is employed at zero temperature \cite{Tselyaev1989,KamerdzhievTertychnyiTselyaev1997,Tselyaev2007},  reduces the number of the time variables accordingly to a single time difference, so that the Fourier transform of Eq. (\ref{bse2}) leads to a single frequency variable equation. The approximation is based on the time projection technique within the Green function formalism, which allows for decoupling of configurations of the lowest complexity beyond $1p1h$ (one-particle-one-hole), such as $1p1h\otimes phonon$ (particle-hole pair coupled to a phonon), from the higher-order ones. Alternatively, one can start directly with two-time response function and arrive to the same result \cite{LitvinovaSchuck2019}. 

Following the time blocking formalism, one notices that the time projection operator introduced at $T=0$ \cite{Tselyaev1989} is not applicable to the finite-temperature case. A generalization found in Refs. \cite{WibowoLitvinova2019,LitvinovaWibowo2018} shows that at $T>0$ the projection operator 
\bea
{\Theta}(14,23;T) &=& \delta_{\sigma_1,-\sigma_2}\theta_{12}(T)\theta(\sigma_1t_{14})\theta(\sigma_1t_{23}),\nonumber \\
\theta_{12}(T) &=& n(\sigma_1\varepsilon_2,T)\theta(\sigma_1 t_{12}) + n(-\sigma_1\varepsilon_1,T)\theta(-\sigma_1 t_{12}),\nonumber \\
\sigma_k &=& \pm 1,
\eea
being introduced into the integral part of Eq. (\ref{bse2}), performs an analogous reduction to one-frequency Matsubara variable. 
Compared to the $T=0$ case, it contains an extra $\theta_{12}(T)$ factor with smooth temperature-dependent Fermi-Dirac distributions, which become the sharp Heaviside functions when recovering the $T\to 0$ limit, so that $\theta_{12}(T) \to 1$. Thereby, the Fourier image of Eq. (\ref{bse2})
\bea
{\cal R}_{np',pn'}(\omega,T) = \tilde{\cal R}_{np}(\omega,T) \delta_{pp'} \delta_{nn'} + \nonumber\\
+ \tilde{\cal R}_{np}(\omega,T) \sum\limits_{p^{\prime\prime}n^{\prime\prime}} 
{\tilde{\cal W}}_{np^{\prime\prime},pn^{\prime\prime}}(\omega,T) 
{\cal R}_{n^{\prime\prime}p',p^{\prime\prime}n'}(\omega,T)\nonumber \\
\label{bse3}
\eea
has the form of the Dyson equation with one energy, or frequency, $\omega$ transferred to the system. In Eq. (\ref{bse3}) we imply that the transitions occur between the states of different isospin, i.e. between neutrons ($n,n',n''$) and protons ($p,p',p''$) and replace the number indices by the corresponding letters.
The uncorrelated proton-neutron propagator $\tilde{\cal R}(\omega,T)$ reads:
\begin{equation}
\tilde{\cal R}_{np}(\omega,T) = \frac{n_{pn}(T)}{\omega - \varepsilon_{n} + \varepsilon_{p} }, 
\label{resp0}
\end{equation}
where $n_{pn}(T) = n_{p}(T )- n_{n}(T)$  and ${\tilde{\cal W}}(\omega,T)$ is the interaction amplitude: 
\be
{\tilde{\cal W}}_{np',pn'}(\omega,T) = \tilde{V}%
_{np',pn'}(T)
+ \Phi_{np',pn'}(\omega,T).
\label{W-omega-tba}%
\ee
In the isospin-flip, or charge-exchange, channels the static part of the interaction $\tilde{V}$ is represented by the exchange of $\pi$ and $\rho$ mesons carrying isospin and the Landau-Migdal term ${\tilde V}_{\delta\pi}$, which ensures the correct short-range behavior:
\be
{\tilde V} =  {\tilde V}_{\rho} + {\tilde V}_{\pi} + {\tilde V}_{\delta\pi}. 
\label{mexch}
\ee
The $\rho$-meson is parametrized according to Ref. \cite{Lalazissis1997}, the pion-exchange is treated as in a free space, and the strength of the last term is adjusted to the GT response of $^{208}$Pb \cite{PaarNiksicVretenarEtAl2004a}, in the absence of the explicit Fock term \cite{LongVanGiaiMeng2006,LiangVanGiaiMengEtAl2008,LiangZhaoRingEtAl2012}.
The amplitude $\Phi(\omega,T)$ comprises all the dynamical effects of PVC:
\begin{eqnarray}%
\Phi_{np',pn'}^{(ph)}(\omega,T)  = \frac{1}{n_{p'n'}(T) } 
\sum\limits_{p''n''\mu} \sum\limits_{\eta_{\mu}=\pm1}
\eta_{\mu} \xi^{\mu\eta_{\mu};n''p''}_{np,n'p'}  
\nonumber \\
\times\frac{ \bigl(N(\eta_{\mu}\Omega_{\mu}) + n_{p''}(T)\bigr)\bigl(n(\varepsilon_{p''}-\eta_{\mu}\Omega_{\mu},T) - n_{n''}(T)\bigr)}{\omega-\varepsilon_{n''}+\varepsilon_{p''}-\eta_{\mu}\Omega_{\mu} }. 
\nonumber\\
\label{phiph}%
\end{eqnarray} 

The phonon vertex matrices $\zeta^{\mu\eta_{\mu}}$
\bea
\xi^{\mu\eta_{\mu};56}_{12,34} &=& \zeta^{\mu\eta_{\mu}}_{12,56}\zeta^{\mu\eta_{\mu}\ast}_{34,56}, \ \ \ \ \ \ 
\zeta^{\mu\eta_{\mu}}_{12,56} = \delta_{15}\gamma^{\eta_{\mu}}_{\mu;62} - \gamma^{\eta_{\mu}}_{\mu;15}\delta_{62},\nonumber\\
\gamma_{\mu;13}^{\eta_{\mu}} &=& \delta_{\eta_{\mu},+1}\gamma_{\mu;13} + \delta_{\eta_{\mu},-1}\gamma_{\mu;31}^{\ast}
\label{phonons}
\eea
and the phonon frequencies $\Omega_{\mu}$ are pre-calculated within the FT-RRPA approach, see Refs. \cite{LitvinovaWibowo2018,WibowoLitvinova2019} for more details.
 The index "$\mu$" in Eqs. (\ref{phiph},\ref{phonons}) includes the full set of phonon quantum numbers, such as angular momentum, parity, and frequency, at the given temperature. 
 The new entities in the numerators of Eq. (\ref{phiph})
$N(\Omega) = 1/ (e^{\Omega /T}-1)$ in Eq. (\ref{phiph}) are bosonic occupation factors associated with the phonons emitted and absorbed in the intermediate states of the proton-neutron pair propagation in the PVC picture.

{The response to a specific external field is associated with the spectral function $S(\omega)$ and transition probabilities $B_{\nu}$ as:
\be
S (\omega) = -\frac{1}{\pi} \lim\limits_{\Delta \to 0} \text{Im} \Pi(\omega + i\Delta) = 
\sum\limits_{\nu} B_{\nu}\delta(\omega - \omega_{\nu}). 
\label{specfun}
\ee
expressed via the polarizability $\Pi(\omega)$ 
\be
\Pi (\omega + i\Delta) = \langle V^{(0)} {\cal R} V^{(0)\dagger}\rangle = 
\sum\limits_{\nu} \frac{B_{\nu}}{\omega - \omega_{\nu} + i\Delta}. 
\label{polarizability}
\ee
In this work we focus on the raising Gamow-Teller (GT$_+$) operator as an external field:
\be
V^{(0)}_{GT_+} = \sum\limits_{i=1}^A \Sigma(i)\tau_+(i),
\label{extfield}
\ee
 where $\Sigma$ is the relativistic spin operator.  
To be compared with the observed spectra of excitations, the spectral function ${S}(\omega)$ should be corrected by an additional factor due to the detailed balance \cite{WibowoLitvinova2019,Dzhioev2015}: 
\be
 \label{Strength}
\tilde{S}(\omega)=\frac{S(\omega)}{1-e^{-(\omega-\delta_{np})/T}},
\ee
where $\delta_{np} = \lambda_{np} + M_{np}, \lambda_{np} =  \lambda_{n} -  \lambda_{p}$  is the difference between
neutron and proton chemical potentials in the parent nucleus and $M_{np} = 1.293$ MeV is the neutron-proton mass
splitting. In our implementation, the response to the conjugate operators $\tau_{\pm}$ is computed in one procedure, so that the resulting spectral functions are located at positive and negative frequencies. Therefore, the reference energy  $\delta_{np}$ is determined as a boundary between them. Obviously, at $T = 0$ the spectral function and the strength function coincide.

As the functions $S(\omega)$ and ${\tilde S}(\omega)$ are formally singular, for representation purposes a finite value of the imaginary part of the energy variable (smearing parameter) $\Delta$ is used. It provides a smooth strength distribution, because the Dirac delta-function is difficult to visualize, but preserves the integrals under the spectral peaks as well as the physical values of the transition probabilities, which do not depend on the value of $\Delta$. The denominator
 in Eq. (\ref{Strength}) is important only for the excitation energies $|\omega-\delta_{np}| \leq T$, otherwise it is mostly close to unity. It is nearly negligible for the general  features of the strength distribution, however, it is taken into account in the calculations of the electron capture rates.

%
\begin{figure}
\begin{center}
\includegraphics[scale=0.45]{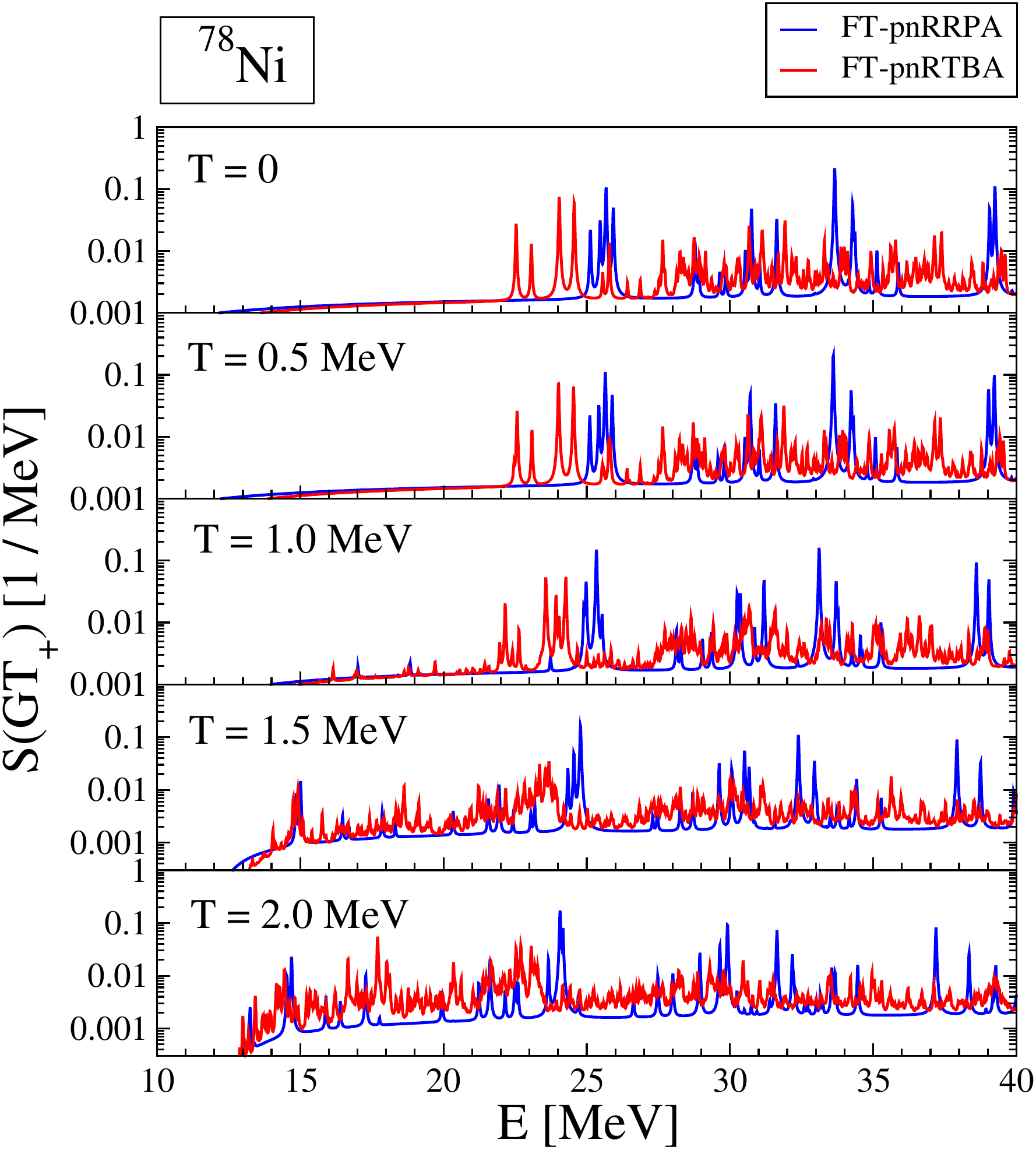}
\end{center}
\caption{Evolution of the Gamow-Teller GT$_+$ spectrum in $^{78}$Ni with temperature within FT-pnRRPA and FT-pnRTBA. }
\label{78ni}%
\end{figure}
%

\section{Details of calculations, results and discussion}
\label{Results}

The FT-pnRTBA model described in Section \ref{Formalism} was applied to calculations of the GT$_+$ strength in the neutron-rich doubly-magic nucleus $^{78}$Ni and the neighboring even-even nuclei $^{76,80}$Ni, $^{76}$Fe and $^{80}$Zn.  As in the previous application of the formalism to beta decay \cite{LitvinovaRobinWibowo2020}, on the first step we solve the set of the relativistic mean field (RMF) equations with the NL3 parametrization, which implies a self-interaction in the scalar sigma-meson sector \cite{Lalazissis1997}.  The thermal occupancies of Eq. (\ref{FermiDirac}) are introduced into the fermionic densities which participate in the self-consistent RMF cycle. The set of the obtained temperature-dependent single-particle Dirac spinors and the corresponding single-nucleon energies form the basis for further calculations. 
The second step consists of solving the finite-temperature relativistic random phase approximation (FT-RRPA) equations and obtaining the phonon vertices $g^{m}$ and their frequencies $\omega_{m}$. 
The set of the resulting phonons, 
together with the thermal RMF single-nucleon basis, forms the $ph\otimes$phonon configurations for the PVC amplitude of Eq. (\ref{phiph}).
As the third step,  Eq. (\ref{bse2}) contracted with the GT$_+$ operator, is solved, the solution is contracted with the second GT$_+$ operator and the spectral function is extracted according to Eq. (\ref{specfun}). 
The particle-hole ($ph$) configurations with the energies $\varepsilon_{ph}\leq100$ MeV and the antiparticle-hole ($\alpha h$) ones with $\varepsilon_{\alpha h}\geq-1800$ MeV were included in the FT-RRPA calculations for the vibrational spectrum, that provides an acceptable convergence of the results. The phonons with the quantum numbers of spin and parity $J^{\pi}=2^{+},\;3^{-},\;4^{+},\;5^{-},\;6^{+}$ below the energy cutoff of 20 MeV and with the reduced transition probabilities $B(EL)$ equal or more than 5\% of the maximal one (for each $J^{\pi}$) comprised the phonon model space, for all temperatures.  Another truncation was made on the 
single-particle intermediate states $n'', p''$ in the summation of Eq. (\ref{phiph}): only the phonon matrix elements with the energy differences  $|\varepsilon_{p(n)}- \varepsilon_{p''(n'')}| \leq$ 50 MeV were included in the summation. All these truncations are justified by our preceding calculations. The value $\Delta = 0.02$ MeV is adopted for the smearing parameter. This value is sufficiently small to resolve the fine features of the spectral functions and to provide a reliable extraction of the electron capture rates. 

\begin{figure}
\begin{center}
\includegraphics[scale=0.45]{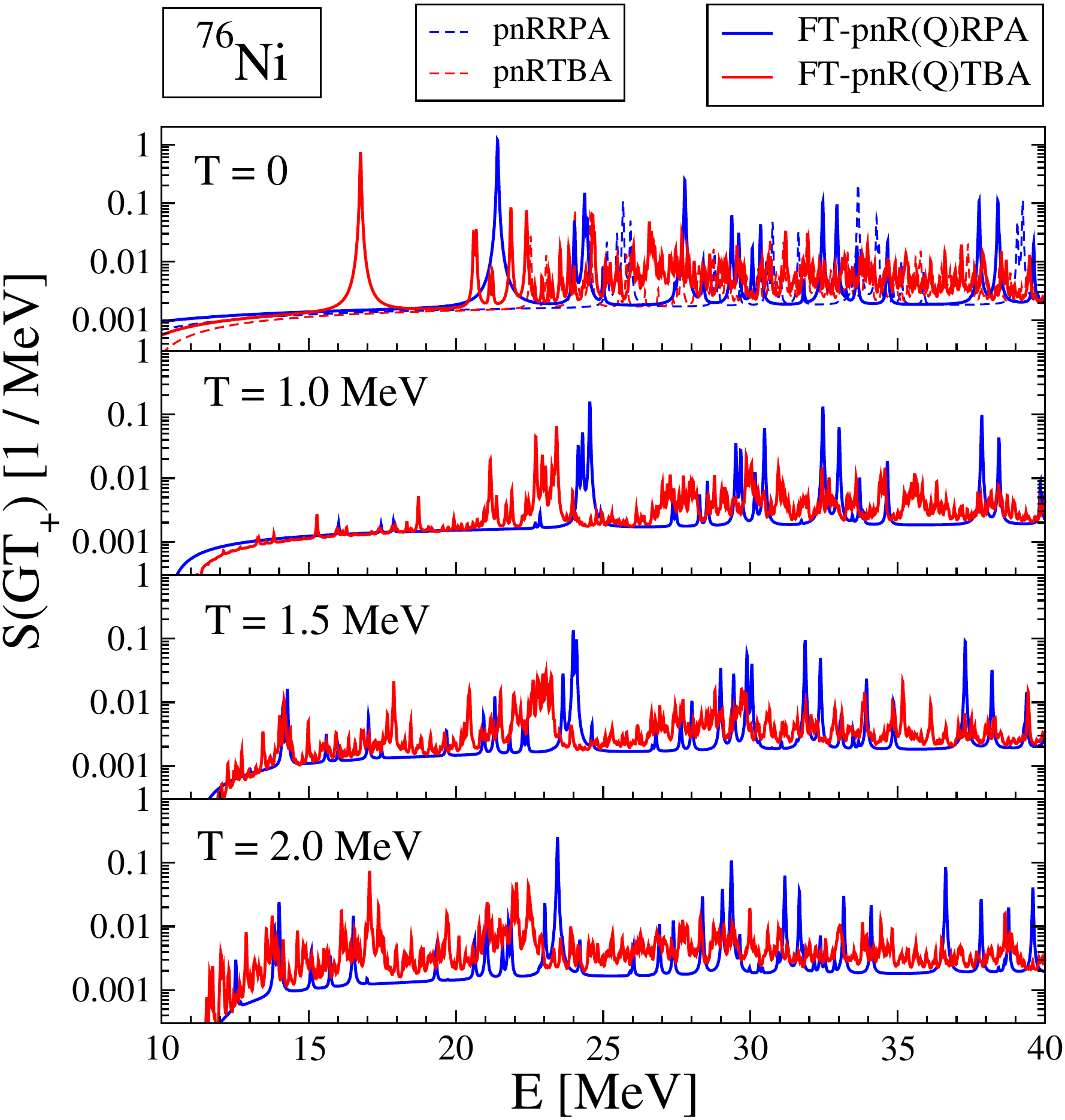}
\end{center}
\caption{Evolution of the Gamow-Teller GT$_+$ spectrum in $^{76}$Ni with temperature. The legends "FT-pnR(Q)RPA" and "FT-pnR(Q)TBA" denote calculations with the superfluid pairing at $T = 0$ and without it at $T \geq 1$ MeV, when the BCS superfluidity vanishes.  
The upper panel shows a comparison of the approaches with (FT-pnRQRPA, FT-pnRQTBA, solid curves) and without (FT-pnRRPA, FT-pnRTBA, dashed curves) superfluid pairing at $T = 0$.
}
\label{76ni}%
\end{figure}
\begin{figure}
\begin{center}
\includegraphics[scale=0.45]{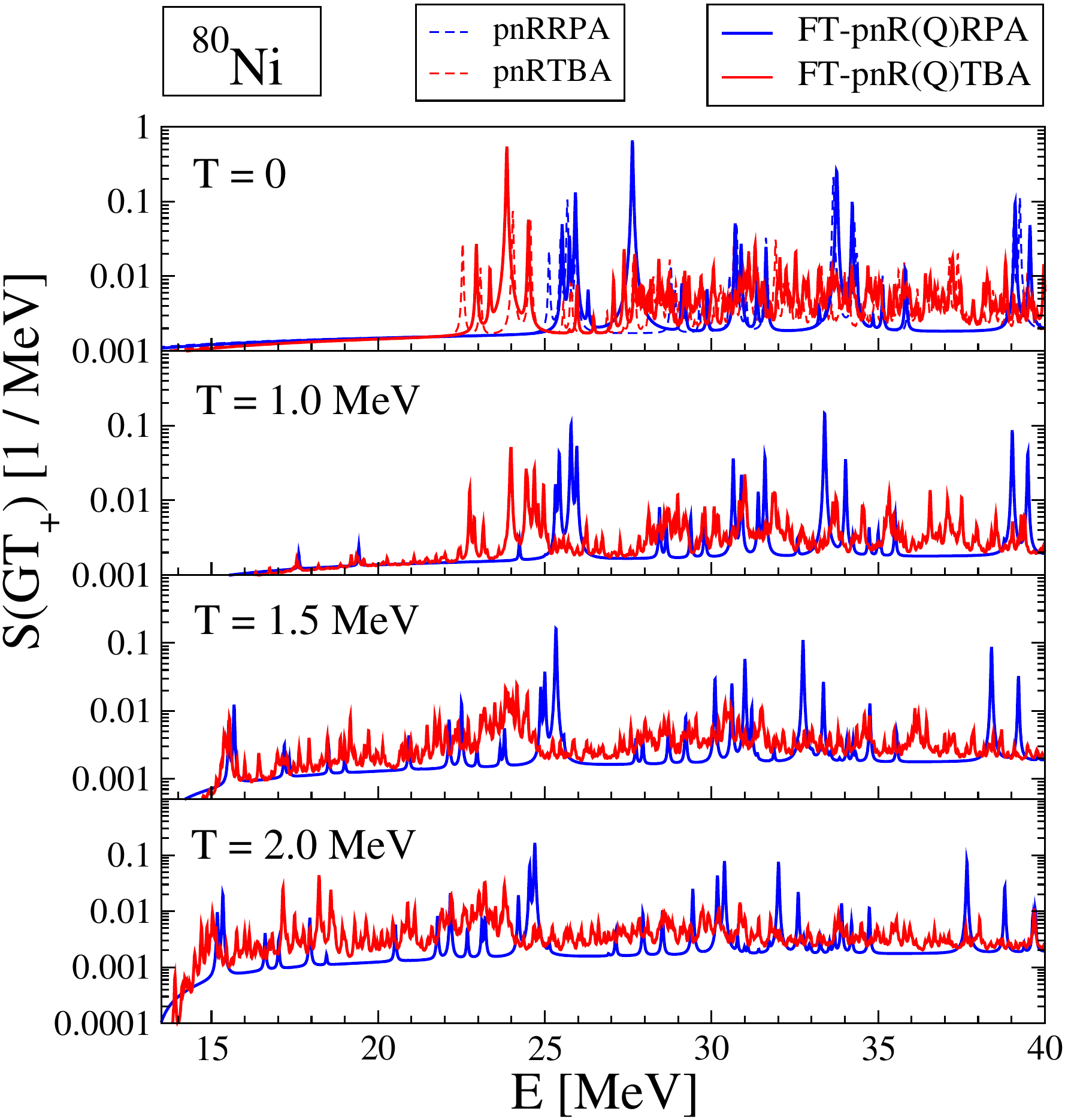}
\end{center}
\caption{Same as in Fig. \ref{76ni}, but for $^{80}$Ni. }
\label{80ni}%
\end{figure}

In the present work we employ a standard Bardeen-Cooper-Schrieffer (BCS) technique to describe nuclear superfluidity.
While $^{78}$Ni is a closed-shell nucleus and, in this framework, does not exhibit superfluid properties, its even-even neighbors do. 
The BCS approximation to nuclear superfluidity also implies that in open-shell nuclear systems superfluid pairing correlations vanish at the critical temperature $T_c \sim 0.6\Delta(0)$, where  $\Delta(0)$ is the pairing gap at $T = 0$. The coefficient between $T_c$ and $\Delta(0)$ may vary from system to system in the relatively narrow limits, in particular, it was found in Ref. \cite {Wibowo2020} that in $^{68}$Ni it takes the value 0.7. We assume that in all isotopes under consideration superfluidity vanishes within $0.5 \leq T\leq 1$ MeV interval. Moreover, the corresponding phase transition is quite sharp, in particular, at temperatures $0 \leq T\leq 0.5$ MeV there is almost no change in the single-particle properties, as we have verified in Ref. \cite{Wibowo2020}. Based on these observations, we assume that in open-shell nuclei under consideration the GT$_+$  spectral functions do not change considerably in the temperature range $0 \leq T\leq 0.5$ MeV. Therefore, at these temperatures we can safely use the zero-temperature spectral functions, which we calculate within the proton-neutron relativistic quasiparticle time blocking approximation (pnRQTBA) \cite{RobinLitvinova2016} fully accounting for superfluid pairing. Starting from the temperature $T = 1$ MeV, the FT-pnRTBA without pairing is then fully legitimate and, if needed, the results in the temperature range $0.5 \leq T\leq 1$ MeV can be interpolated. In this way, we can avoid complications like adopting  pnRQTBA for finite temperatures, which is only needed in a very narrow temperature interval.

The spectral functions computed in this framework are displayed in Figures \ref{78ni} - \ref{80zn}. Fig. \ref{78ni} shows the GT$_+$ transitions in $^{78}$Ni calculated up to high excitation energies within FT-pnRTBA in comparison with the FT-pnRRPA. In our formalism the latter is obtained as a solution of Eq. (\ref{bse3}), if the dynamical PVC term $\Phi(\omega,T)$ in the interaction amplitude is fully neglected. In this way, the difference between the two distributions isolates the effects of the dynamical correlations, which makes their assessment obvious. One can see that at $T = 0$  FT-pnRTBA the spectral function is considerably more fragmented than the FT-pnRRPA one, that is the common feature of the approaches with PVC kernels. Indeed, the dynamical kernel of Eq. (\ref{phiph}) contains the poles of the $ph\otimes phonon$ character, which provide a richer structure of the FT-pnRTBA spectral functions for all types of response. As we have investigated previously \cite{RobinLitvinova2016,RobinLitvinova2018}, typically the positions of the first peaks shift to lower energies when the PVC effects are taken into account in the leading approximation, however, further inclusion of the ground state correlations caused by PVC induces a counter trend, which may shift the lowest peaks upward, although not necessarily to their original placements within the pnR(Q)RPA \cite{Robin2019}.
Moreover, configurations of higher complexity, such as correlated multiparticle-multihole ones, may induce further redistribution of the strength, including the one of the low-energy tails, although the effects of such higher-rank  correlations weaken with their complexity growth. Until now, however, numerical studies are only available for non-isospin-flip excitations up to the correlated three-particle-three-hole ($3p3h$) configurations \cite{LitvinovaSchuck2019}.

As we will see in the following, for the electron capture rates, which are extracted from these spectral functions, the presence of transitions at lowest energies caused solely by the dynamical PVC effects plays a crucial role, especially at low electron densities. With the temperature growth the spectral functions evolve because of the thermal unblocking: as the Fermi-Dirac distribution of Eq. (\ref{FermiDirac}) become more diffuse, more transitions appear with the sizable strength. As a result, similarly to the case of the neutral transitions discussed in Refs. \cite{LitvinovaWibowo2018,WibowoLitvinova2019}, the entire spectral function undergoes stronger fragmentation in both FT-pnRRPA and FT-pnRTBA, while the PVC effects further enhance fragmentation in the latter approach. In particular, the spreading of the states toward lower transition frequencies is the most impactful effect, which has direct implications for the electron capture rates. 
In this study we calculate the temperature evolution of the GT$_+$ strength and the subsequently extracted electron capture rates up to $T = 2$ MeV on the temperature grid with 0.5 MeV step. This choice is determined by the fact that these nuclear temperatures are of the primary importance for supernovae evolution and the temperature dependencies of the quantities under study are rather smooth, so that they can be interpolated, if needed, between the given mesh points. The GT$_+$ strength in $^{78}$Ni computed for $T = 0.5$ MeV comes out almost identical to the one obtained at 
$T = 0$. This result indicates that the temperature of $T = 0.5$ MeV is still too low to induce sizable changes in both thermal mean field and the transition amplitudes of both FT-pnRRPA annd FT-pnRTBA.
We will use this result in the following to justify the use of $T = 0$ GT$_+$ strength for calculations at $T = 0.5$ MeV for open-shell nuclei. 

\begin{figure}
\begin{center}
\includegraphics[scale=0.45]{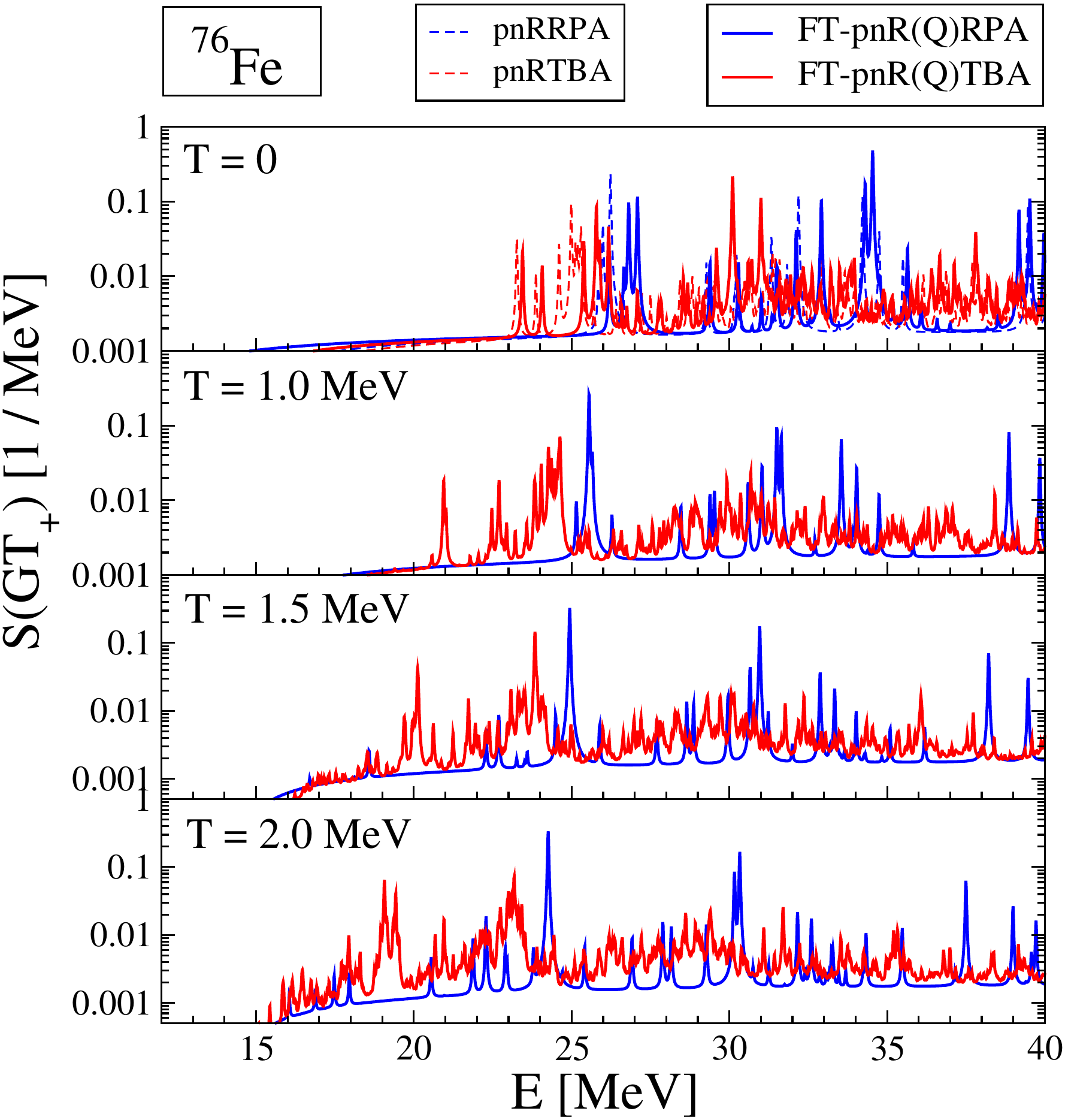}
\end{center}
\caption{Same as in Fig. \ref{76ni}, but for  $^{76}$Fe. }
\label{76fe}%
\end{figure}
\begin{figure}
\begin{center}
\includegraphics[scale=0.45]{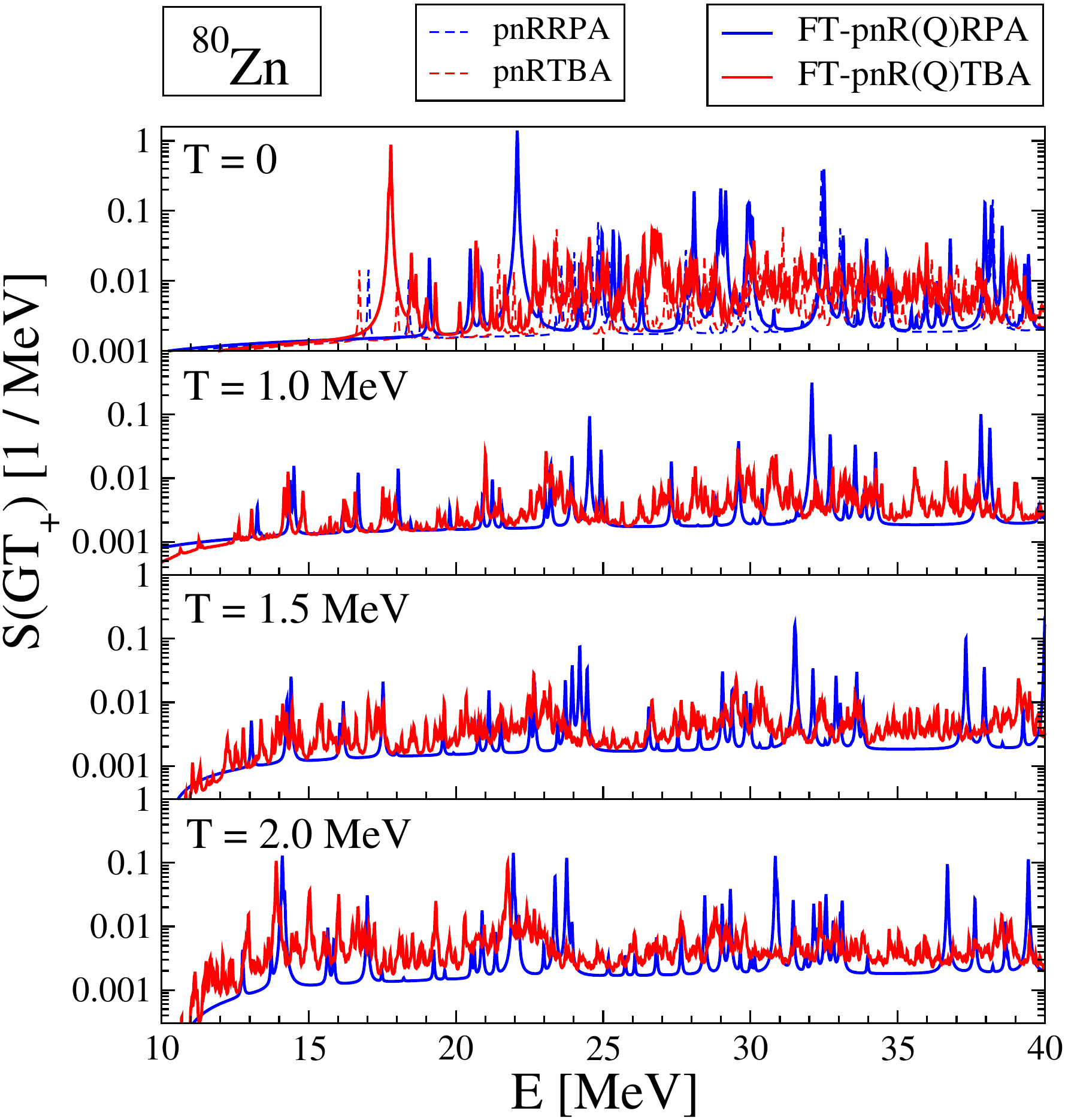}
\end{center}
\caption{Same as in Fig. \ref{76ni}, but for  $^{80}$Zn. }
\label{80zn}%
\end{figure}

At $T = 1$ MeV the picture begins to change. On the third panel of Fig. \ref{78ni} it can be seen that, while the spectral function is still not very different from that at $T = 0$ at high energies, it develops new low-energy transitions which are absent at $T = 0$. They are well seen on the logarithmic scale as the signatures of the thermal unblocking. Its nature becomes clear if one notices that in the lowest approximation, when no interaction between fermions is active besides the mean field (this limit corresponds to ${\tilde{\cal W}}(\omega, T) = 0$), the solutions of the Eq. (\ref{bse3}) is the uncorrelated mean-field propagator ${\tilde{\cal R}}(\omega, T)$ of Eq. (\ref{resp0}). Its numerator is the difference between proton and neutron occupancies which can only take the values of 0 and 1 at $T = 0$. At $T > 0$ the fractional occupancies become possible and, therefore, the transitions take place not only across the Fermi surfaces, but also when both single-nucleon states lie above or below them. With the temperature growth the intensity of such transitions also increases and, thus, more and more of them become visible in the spectra. Switching on the interaction amplitudes causes redistribution and fragmentation of the excited states while keeping the effects of the thermal unblocking. Indeed, the lowest two panels of Fig. \ref{78ni} demonstrate the increasing amount and intensity of the low-energy states while temperature raises to $T = 1.5$ MeV and eventually to $T = 2$ MeV.

Fig. \ref{76ni} illustrates the evolution of the GT$_+$ strength with temperature for the nucleus $^{76}$Ni. In contrast to $^{78}$Ni, this nucleus is open-shell in the neutron subsystem, so that neutron pairing correlations are included into the description at $T = 0$. The upper panel of Fig. \ref{76ni} with the zero-temperature strength distributions displays the results obtained with and without superfluid pairing, both with and without PVC. One can see that, while the high-energy parts of the strength are not much influenced by pairing, the low-energy parts are drastically different, for the calculations both with (pnRQTBA vs pnRTBA) and without (pnRQRPA vs pnRRPA) the dynamical PVC effects. Namely, the superfluid pairing introduces additional possibilities for the low-energy transitions, that is reflected in the resulting spectra as relatively strong low-energy peaks at  21.4 MeV and 16.8 MeV within the pnRQRPA and pnRQTBA, respectively (they are dominated by $pf_{7/2} \to nf_{5/2}$ transition). As it is mentioned above, we assume that at $T = 0.5$ MeV the GT$_+$ strength is not very different from the one at $T = 0$, which is consistent with the result obtained directly for $^{78}$Ni. At $T = 1$ MeV there are two major effects: (i) disappearance of the superfluid pairing that is clear from the disappearance of the strong peaks at low energy and (ii) the formation of the thermally unblocked states in full analogy with the case of $^{78}$Ni. Moreover, for the temperatures $T \geq 1$ MeV, when superfluidity has no influence, the overall spectral pattern of GT$_+$ in $^{76}$Ni is very close to the one of $^{78}$Ni. This result is consistent with the fact that the compositions of these nuclei differ by only two neutrons on predominantly $1g_{9/2}$ orbital, so that, after the transition to the non-superfluid phase, with the temperature increase their Fermi surfaces and the arrangements of the single-particle states around them become very similar. Aa a result, the excitation spectra look similar, too.

Figs. \ref{80ni} - \ref{80zn} display the temperature evolution of the GT$_+$ strength for $^{80}$Ni, $^{76}$Fe and $^{80}$Zn, respectively, in the same manner as Fig. \ref{76ni}. While the low-energy spectra at $T =0$ differ remarkably as a consequence of the particular rearrangements of the shell structure near the Fermi surfaces of these systems, after the transition to the non-superfluid phase the general pattern and the evolution of the spectra with temperature exhibit the same features. One may notice that the pair of nuclei $^{76}$Ni and $^{80}$Zn as well as the pair $^{76}$Fe and $^{80}$Ni show similarities in the structure their low-energy spectra, that may be related to the fact that each pair shares the same value of isospin.

Based on the calculated GT$_+$ strength, we extracted the electron capture (EC) rates at some typical thermodynamical conditions that occur during the star evolution. In this first application of our approach to the EC rates we do not focus on the accuracy of their determination, but rather use the simplest prescription in order to understand, how the effects of complex many-body correlations propagate to the EC process in stellar environments. Thus, the EC rates were calculated in the zero momentum transfer limit, when only the GT$_+$ transitions contribute.  The impact of forbidden transitions on EC rates for nuclei with $A \sim 80-90$, such as $^{78}$Ni, $^{82}$Ge, $^{86}$Kr and $^{88}$Sr within QRPA approaches was analyzed in Ref. \cite{Dzhioev2020}. It was shown, in particular, that at temperatures around 10 GK the forbidden transitions start to play a noticeable role at high densities lg$(\rho Y_e) \equiv \text{log}_{10}(\rho Y_e)  \approx 11$. Within our approach, the contribution of the first-forbidden transitions to the beta decay rates was investigated quantitatively and discussed for the r-process waiting-point nuclei $^{78}$Ni and $^{132}$Sn in Ref. \cite{LitvinovaRobinWibowo2020}. It was found that in the FT-pnRTBA approach, which reproduces successfully the observed beta decay rates, their contributions are relatively minor at zero temperature (in particular, we found 6\% for $^{78}$Ni and 20\% for $^{132}$Sn), but can increase with the temperature growth (up to 40\% and 55\% at T = 2 MeV, respectively). These numbers, however, are very sensitive to fine details of the calculated spectra and to the procedure of extracting the rates. For instance, the importance of the forbidden transitions is often emphasized in the literature, where the calculations are performed within the (thermal) (Q)RPA, however, it is not clear how reliable those estimates are as this approach can not reproduce the observed fine structure of the spin-isospin response without introducing artificial terms in the residual interaction with adjustable parameters. 

Based on the results of Ref. \cite{LitvinovaRobinWibowo2020} for the beta decay, we expect a similar amount of contribution from the first-forbidden transitions to the EC rates.  In the zero momentum transfer limit the common prescriptions give \cite{Langanke2001,Dzhioev2010}:
\bea
\lambda_{A,Z}(\mu_e,T) &=& \frac{g_A^2\ \text{ln2}}{K} \int\limits_{1}^{\infty}d\varepsilon_e \varepsilon_e p_e f(\varepsilon_e,\mu_e,T) F(A,Z,\varepsilon_e)\times\nonumber\\
&\times&\int\limits_{-\infty}^{\varepsilon_e}dE(\varepsilon_e - E)^2S_{GT_+}(E,T),
\label{rate}
\eea
where energies are in the units of electron mass $m_e$, $p_e^2 = \varepsilon_e^2 -1$, and $F(A,Z,\varepsilon_e)$ is the Fermi function \cite{Langanke2001}. The values $K = 6163.4$ s and $g_A = 1.27$ are adopted for the pre-integral constants.

The reference approximation to the EC rates in neutron-rich nuclei of the $pf-sdg$ shells, which is widely used in astrophysical simulations, is the parameterization of Ref. \cite{Langanke2003} based on the analytical approach of Ref.  \cite{Fuller1985}. It has the form:
\be
\lambda = \frac{B\ \text{ln2}}{K}\Bigl(\frac{T}{m_ec^2}\Bigr)^5[F_4(\eta) - 2\chi F_3(\eta) + \chi^2F_2(\eta)]
\label{param}
\ee
containing the Fermi integrals $F_k(\eta)$, such as 
\be
F_k(\eta) = \int\limits_0^{\infty} \frac{d{\varepsilon}\ \varepsilon^k}{1 + \text{exp}(\varepsilon - \eta)}
\ee
with $\chi = -(Q+\Delta E)/T$, $\eta = \chi + \mu_e/T$  and the transition strength $B$ and the transition energy $\Delta E$ as fitted parameters. 
The value of the electron density, which is typically used in combination with the electron-to-barion ratio $\rho Y_e$ determines the electron chemical potential, which is found by solving the equation \cite{Langanke2001}:
\be
\rho Y_e(T) = \frac{1}{\pi^2N_A}\Bigl(\frac{m_ec}{\hbar}\Bigr)^3
\int\limits_{0}^{\infty}\Delta S_{ep}(\varepsilon_e,\mu_e,T)p_e^2 dp_e
\label{mue}
\ee
with $\Delta S_{ep} = S_e - S_p$, the Fermi-Dirac distribution for electrons $S_e$
\be
S_e(\varepsilon_e,\mu_e,T)  = \frac{1}{1 + \text{exp}\Bigl(\frac{\varepsilon_e - \mu_e}{T}\Bigr)}
\ee
and $S_p(\varepsilon,\mu,T) = S_e(\varepsilon,-\mu,T)$ for positrons.
As the parameters $B$ and $\Delta E$ have been determined in Ref. \cite{Langanke2003} from fitting the EC rates computed microscopically within the combined shell-model Monte Carlo and RPA approach \cite{Langanke2001a} at high electron densities, a comparison to the parameterization of Eq. (\ref{param}) can serve as a comparison to the latter approach in these density regimes.
Note that under this condition the EC rates of Eq. (\ref{param}) are fully determined by the Q-values $Q = M_f - M_i$ being the mass difference between the final and initial nuclei.
In our calculations the Q-values were taken from Ref. \cite{Moeller2016}.
\begin{figure}
\begin{center}
\includegraphics[scale=0.35]{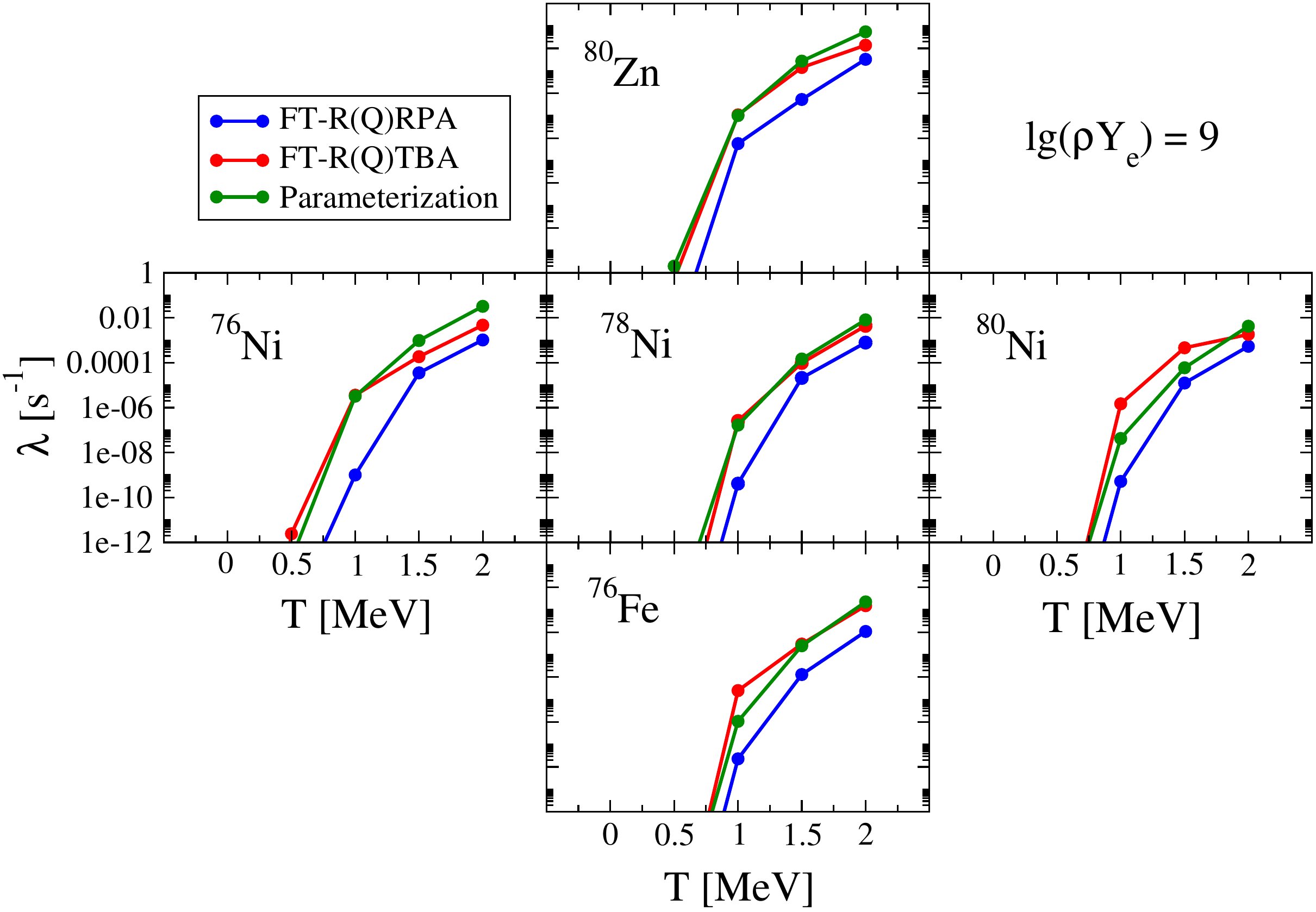}
\end{center}
\caption{Electron capture rate as a function of temperature for neutron-rich even-even nuclei around $^{78}$Ni at lg$(\rho Y_e)$ = 9. }
\label{rates9}%
\end{figure}

The EC rates extracted from the FT-pnR(Q)RPA and FT-pnR(Q)TBA GT$_+$ spectra for the electron density  $\text{lg}(\rho Y_e)$ = 9 according to Eq. (\ref{rate}) are displayed in Fig. \ref{rates9} as functions of temperature, in comparison to those of Eq. (\ref{param}). From the structure of Eq. (\ref{rate}) and 
the general kinematical condition of the EC process \cite{Langanke2001,Titus2018,Ravlic2020} it can be established that the electron chemical potential or, more precisely, the value of $\mu_e - M_{np}$ plays the role of a diffuse threshold for the nuclear transitions: only the transitions with the energies around and below this value contribute to the rates considerably, while those above it are exponentially suppressed. Furthermore, $\mu_e$ is uniquely defined by  Eq. (\ref{mue}), i.e. by the given $\rho Y_e$ value and depends only weakly on the temperature within the $0 \leq T \leq 2$ MeV range. For instance, the density $\text{lg}(\rho Y_e)$ = 9 corresponds to $5.17 \geq \mu_e \geq 2.88$ MeV, below which the nuclei under consideration do not exhibit GT$_+$ transitions. Therefore, for $\text{lg}(\rho Y_e)$ = 9 we have obtained very low EC rates: from almost zero at $T = 0$ to $\sim~10^{-3}-10^{-2}$ s$^{-1}$ at $T = 2$ MeV. However, even in this situation one can see from Fig. \ref{rates9} that the EC rates extracted from the FT-pnR(Q)RPA strength distributions are substantially smaller than those obtained from FT-pnR(Q)TBA ones. The difference between them is larger at lower temperatures and can reach few orders of magnitude. Fig. \ref{rates9} also shows a  good agreement of our FT-pnR(Q)RPA EC rates with those of the parameterization of Eq. (\ref{param}). We conclude that at such relatively low electron densities the complex nuclear correlations beyond those included in the RPA type of approaches are of great importance. The PVC correlations taken into account in FT-pnR(Q)RPA of this work and the shell-model occupancies adopted in RPA of Refs. \cite{Langanke2001a,Langanke2003} and fitted by Eq. (\ref{param}) result in comparable EC rates in nuclei around $^{78}$Ni at $\text{lg}(\rho Y_e)$ = 9.
\begin{figure}
\begin{center}
\includegraphics[scale=0.35]{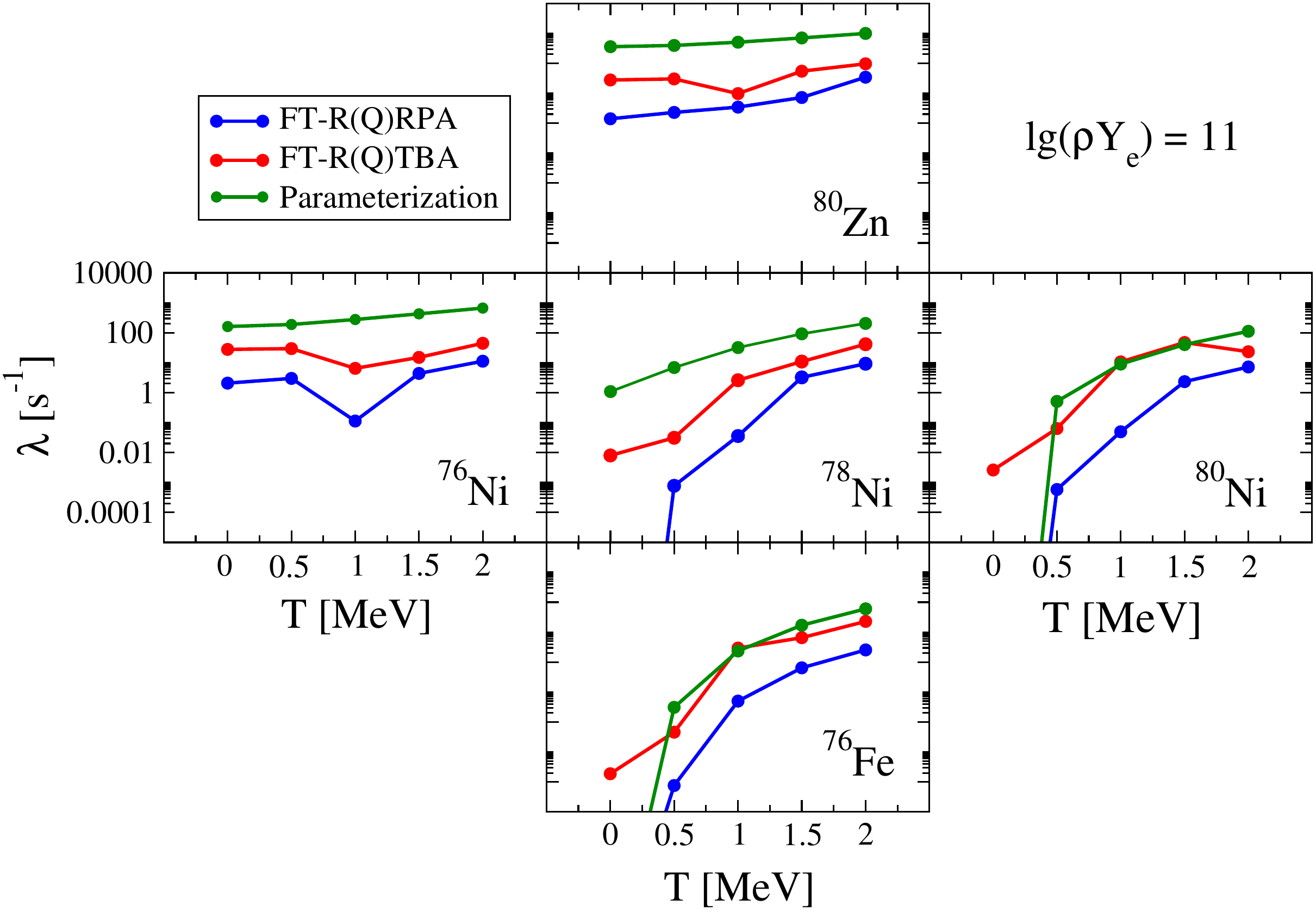}
\end{center}
\caption{Same as in Fig. \ref{rates9}, but at  lg$(\rho Y_e)$ = 11. }
\label{rates11}%
\end{figure}

At higher electron density, such as $\text{lg}(\rho Y_e)$ = 11, the energy window of allowed nuclear Q-values extends to $23.89 \geq \mu_e \geq 23.34$  MeV (with the additional correction for $M_{np}$) for our range of temperatures $0 \leq T \leq 2$ MeV. At this density we have two kinds of situations at $T = 0$: (i) both FT-pnR(Q)RPA  and FT-pnR(Q)TBA generate transitions below or around the threshold energy and (ii) such transitions are only possible in FT-pnR(Q)TBA while the lowest FT-pnR(Q)RPA states are noticeably higher. The situation (i) realizes in $^{76}$Ni and $^{80}$Zn, while the situation (ii) is observed in the rest of the considered nuclei. Respectively, as it is shown in Fig. \ref{rates11}, at low temperatures the influence of the PVC correlations on the EC rates in the second group of nuclei is very big, while for the first group it is more moderate, although also significant as the EC rates still differ by a couple of orders of magnitude in  calculations with and without PVC. With the temperature increase the rates appear  less sensitive to the PVC correlations as the results obtained within FT-pnR(Q)RPA  and FT-pnR(Q)TBA get closer to each other. As we mentioned above, in our description  the open-shell nuclei $^{76}$Ni, $^{80}$Ni, $^{76}$Fe and $^{80}$Zn experience a phase transition to the non-superfluid phase within the $0.5 \leq T \leq 1$ MeV temperature range, that leads to the decrease of the strength amount at the lowest energies. At the same time, thermally unblocked transitions show up in this energy region, that increase the amount of strength. These two counter trends interplay differently in different nuclei, depending on the particular arrangements of the single-particle states  around the Fermi energy. As a result, one can see a non-smooth behavior of the EC rates around $T = 1$ MeV in the nuclei of the first group, while no such effect is visible in nuclei of the second group. Another observation from Fig. \ref{rates11} is that at higher densities the FT-pnR(Q)TBA results are also in a better agreement with the parameterization of Eq. (\ref{param}) than the FT-pnR(Q)RPA ones. Especially good agreement is obtained for the nuclei $^{76}$Fe and $^{80}$Ni at the temperatures $0.5 \leq T \leq 2$ MeV, while it is not so good for the others. Nevertheless, the general trends are reproduced quite reasonably. The significant differences that remain for $^{76,78}$Ni and $^{80}$Zn can be attributed to the use of a different interaction, missing forbidden transitions and still missing correlations of higher-rank in our FT-pnR(Q)TBA approach. However, it is difficult to predict whether these distinctions can increase the EC rates by 1-2 orders of magnitude.  A puzzling discrepancy is found at low temperatures in $^{78}$Ni, where the parameterization of Eq. (\ref{param}) returns quite sizable EC rates of the order of 10 s$^{-1}$, while the model space of our microscopic calculations does not offer possibilities of having transitions that could produce such EC rates. Notice here, that the parameterization of Eq. (\ref{param}) is also known to overestimate the EC rates in neutron-rich nuclei at high densities and temperatures \cite{Raduta2017}. 
A contrasting disagreement also remains for the nuclei $^{76}$Fe and $^{80}$Ni at the temperatures $0 \leq T \leq 0.5$ MeV, where FT-pnR(Q)TBA leads to non-vanishing rates while the parameterization predicts nearly zero rates. 
\begin{figure}
\begin{center}
\includegraphics[scale=0.35]{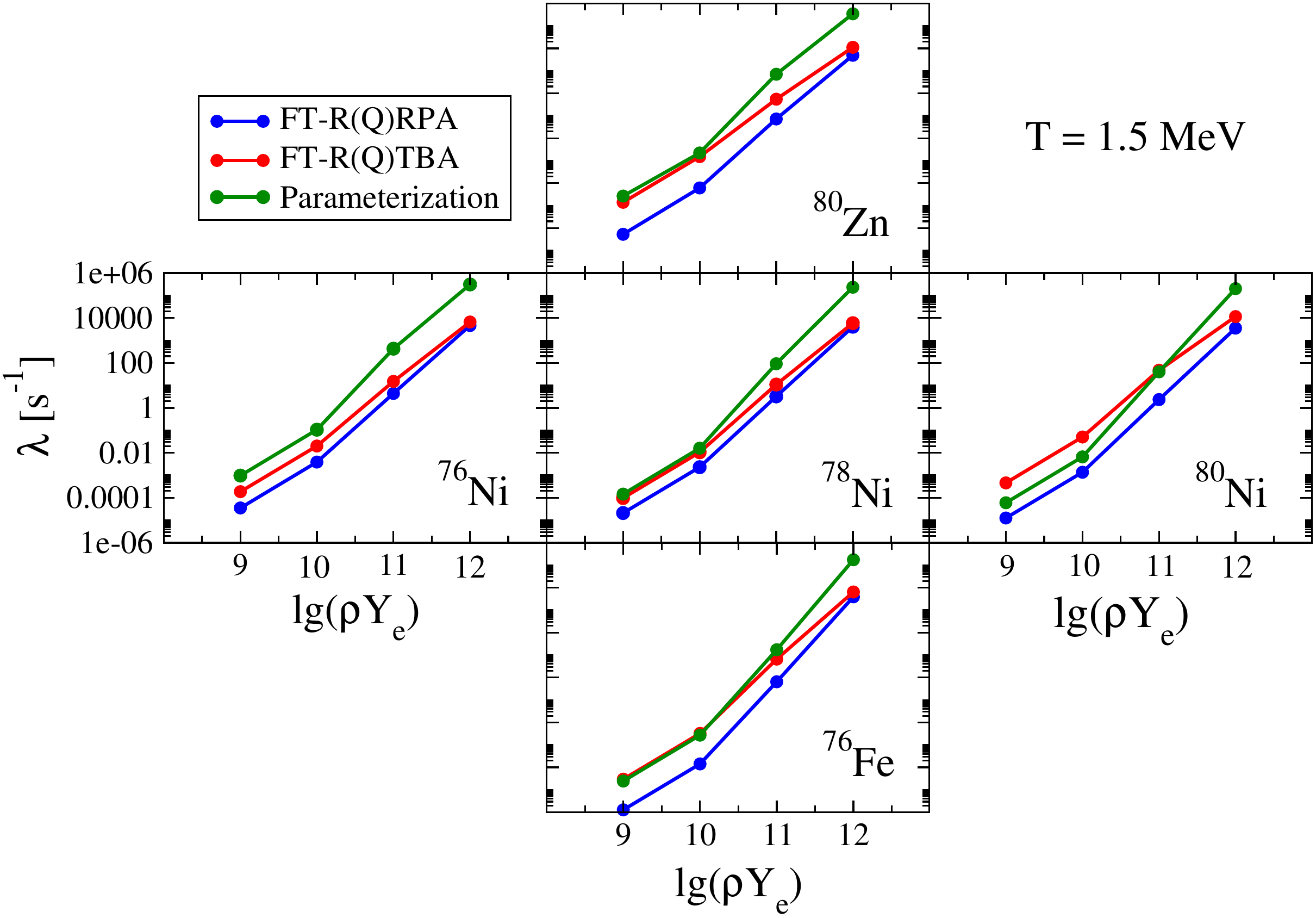}
\end{center}
\caption{Electron capture rate as a function of electron density for neutron-rich even-even nuclei around $^{78}$Ni at $T$ = 1.5 MeV.}
\label{rates1.5}%
\end{figure}
\begin{figure}
\begin{center}
\includegraphics[scale=0.40]{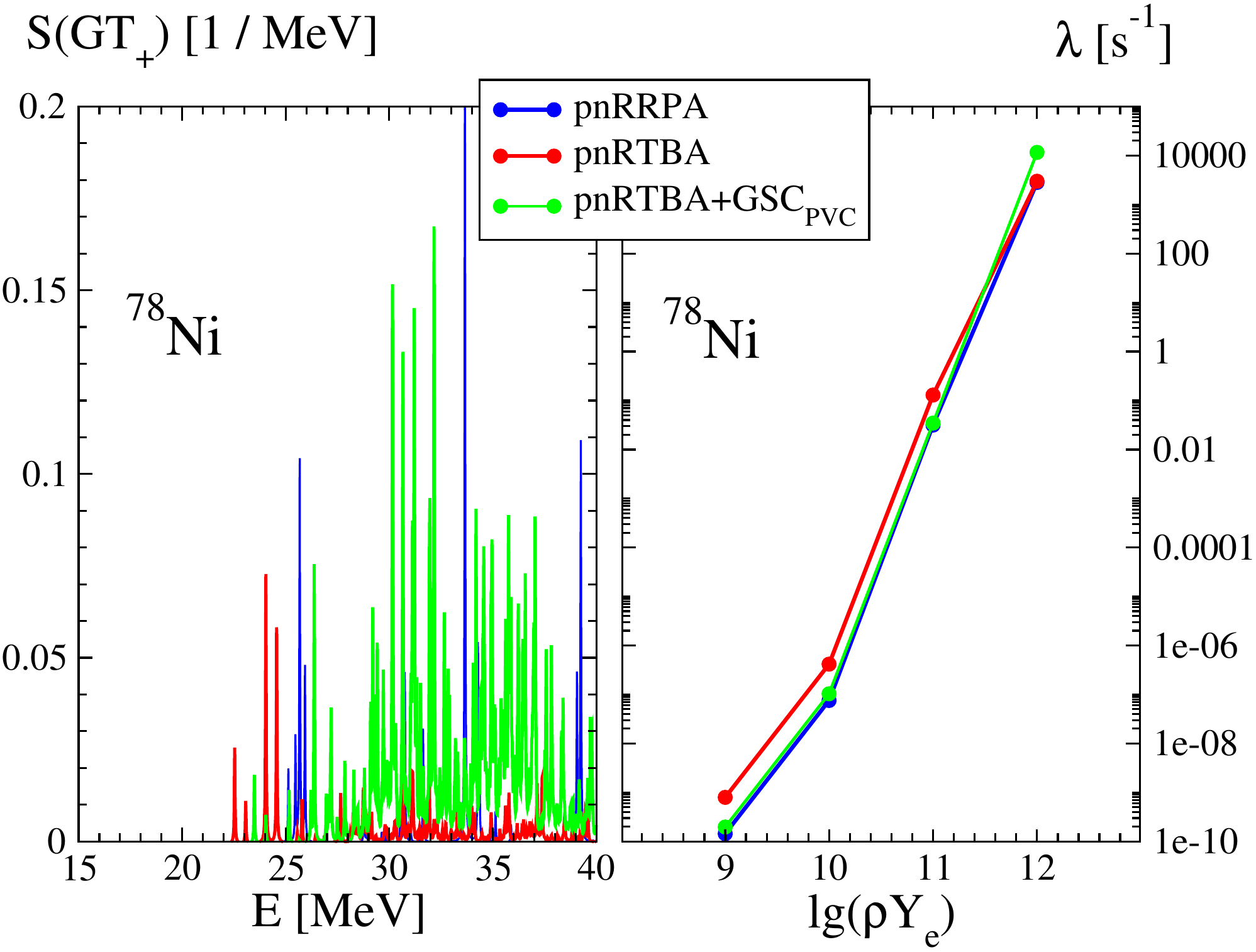}
\end{center}
\caption{Left panel: Gamow-Teller GT$_+$ transitions in $^{78}$Ni computed within pnRRPA and beyond it taking into account the PVC mechanisms: without (pnRTBA, red curve) and with (pnRTBA+GSC$_{\text{PVC}}$, light green curve) ground state correlations induced by PVC. Right panel: electron capture rate at $T = 1$ MeV as a function of electron density for $^{78}$Ni within the respective approximations (temperature dependence of the GT$_+$ strength functions is neglected).}
\label{rates_gsc}%
\end{figure}

The comparison between the three approaches is paraphrased in Fig. \ref{rates1.5} showing the EC rates as functions of the electron densities at fixed temperature $T = 1.5$ MeV. The general trends suggest that (i) the role of PVC correlations included in FT-pnR(Q)TBA levels off at very high densities, however, their importance persists at least up to lg$(\rho Y_e)$ = 11; (ii) FT-pnR(Q)TBA demonstrates generally a better agreement with the parameterization of Ref. \cite{Langanke2003} than FT-pnR(Q)RPA; and (iii) at very high densities the discrepancy between FT-pnR(Q)TBA and the parameterization of Ref. \cite{Langanke2003} increases significantly. It would be interesting to investigate the latter observation further. At high electron densities lg$(\rho Y_e) \approx$ 12, when the electron phase-space factor unlocks the GT$_{+}$ strength up to $\sim$50 MeV, the obtained discrepancy indicates that at such high energies either the GT$_{+}$ spectra are very different in different approaches or the role of forbidden transitions is more important.

In this context, we would like to emphasize again that, indeed, the GT$_{+}$ spectra of neutron-rich nuclei are highly sensitive to the model assumptions and to the complexity of the nuclear wave functions adopted in the theoretical approach. In Ref. \cite{Robin2019}  it was demonstrated how the correlations associated with the time reversed particle-vibration loops, or the ground state correlations caused by PVC (GSC$_{\text{PVC}}$), can unlock GT$_{+}$ transitions in $^{90}$Zr, which are strongly suppressed in both pnRRPA and pnRTBA with the conventional $ph \otimes phonon$ configurations. 
The particularity of the neutron-rich nuclei is that the GT$_{-}$ transitions strongly dominate GT$_{+}$ ones. 
Therefore, due to the conserved Ikeda sum rule, small relative changes in the total strength in the GT$_{-}$ sector corresponds with large relative changes in the total GT$_{+}$ strength. Thus, such GSC$_{\text{PVC}}$ may not appear important for the GT$_{-}$, but play the dominant role for the GT$_{+}$ spectra. Fig. \ref{rates_gsc} illustrates the effect of these complex ground state correlations on the GT$_{+}$ strength in $^{78}$Ni. At the low-energy end of the spectra, one can observe how the PVC effects taken into account within the conventional pnRTBA produce new states at considerably low energies than those of pnRRPA, but with the inclusion of the GSC$_{\text{PVC}}$ this strength is partially pushed back indicating that pnRTBA might overestimate the spreading of the GT$_{+}$ strength to lower energies. Another and more remarkable effect of the  GSC$_{\text{PVC}}$ is seen as the appearance of new states with high intensity in the $30-40$ MeV range. These are the new unlocked transitions which are, in principle, absent in the other, more simple, two approaches. While the formation of this type of strength is explained in detail in Ref. \cite{Robin2019}, here we concentrate on the impact of these correlations on the EC rates. It is illustrated in the right panel of Fig. \ref{rates_gsc}, where we show the rates extracted from the strength functions of the left panel keeping  $T = 1$ MeV for the electron kinematics, in order to better illuminate the effects of correlations.  One can notice that at low electron densities the correlations of the GSC$_{\text{PVC}}$ type lead to lower rates than pnRTBA, almost coinciding with the pnRRPA results. Once the density raises and higher-energy transitions start to contribute, pnRTBA+GSC$_{\text{PVC}}$ competes with pnRTBA and after $\text{lg}(\rho Y_e) \approx$ 11.5 begins to produce higher EC rates.  Since our pnRTBA+GSC$_{\text{PVC}}$ approach does not yet include thermal effects, we do not compare the rates of Fig. \ref{rates_gsc} with the parameterized rates,  but conclude that correlations of the GSC$_{\text{PVC}}$ type have the potential of enhancing the EC rates at high electron densities.

Finally, it should be noted that the recent Refs. \cite{Dzhioev2019,Dzhioev2020} also investigated the EC rates in $^{78}$Ni. This has been done within the formalism of thermo field dynamics confined by the RPA type wave functions. The numerical implementation was based on the Skyrme-Landau-Migdal interaction without thermal modification of the nuclear mean field. Although the authors included the effects of first-forbidden transitions on the EC rates in their study, the GT$_{+}$ spectra and their contributions to the EC rates were presented separately, that allows us to make a meaningful comparison with our case. Since the calculations of Refs. \cite{Dzhioev2019,Dzhioev2020} are based on the RPA type of approximation, they should correspond to our FT-pnRRPA with minimal amount of correlations. Indeed, qualitatively similarly to Refs. \cite{Dzhioev2019,Dzhioev2020}, in FT-pnRRPA with the temperature increase we obtain some general thermal unblocking that leads to the appearance of new transitions at the low end of the spectrum, which begin to be visible at $T = 1$ MeV in Fig. \ref{78ni}. However, we did not observe the strong low-energy $pf_{7/2} \to nf_{5/2}$ transition in the GT$_{+}$ branch at this temperature value, which leads us to somewhat lower EC rates than those of Refs.  \cite{Dzhioev2019,Dzhioev2020}. 



\section{Summary and outlook}
\label{Summary}

In this work we investigated the role of complex nuclear correlations in the stellar electron capture process for the nuclei around $^{78}$Ni which are abundantly produced, for instance, during the stellar collapse at temperatures $T \sim$ 10 GK and densities $\text{lg}(\rho Y_e) \approx$ 11. A more advanced approach (FT-pnRTBA), taking into account complex nuclear correlations originated from coupling between the single-particle and collective degrees of freedom, the particle-vibration coupling,  was compared with the simpler approach (FT-pnRRPA) neglecting these correlations. While the finite-temperature RPA is known conceptually since many decades, the finite-temperature relativistic time blocking approximation was developed only recently. Both approaches adopted for charge-changing transitions on the base of the relativistic effective meson-exchange Lagrangian were applied to calculations of the Gamow-Teller excitations in the $\beta^+$ branch in even-even nuclei around $^{78}$Ni. The electron capture rates for the range of temperatures and electron densities around those of the most abundant production of such nuclei in stars were extracted in the zero momentum transfer limit. 
The GT$_+$ strength distributions and EC rates obtained within these two models were compared to reveal how the PVC correlations, being purely microscopical effects of internal nuclear structure, propagate to the electron capture processes in the stellar media, which determine large-scale features of star evolution. In general, we found that the PVC correlations increase the EC rates. As a consequence, they further reduce the electron-to-baryon ratio leading to lower pressure, thus promoting the gravitational collapse. The concurrent increase of the neutrino flux intensifies the effective cooling that, in turn, allows heavy nuclei to survive the collapse.

The EC rates calculated within the FT-pnRTBA were compared to the existing systematics based on the parameterization of SMMC and RPA calculations and found to be in a partial agreement with that systematics, while the FT-pnRRPA results showed considerably lower EC rates. The overall agreement between the FT-pnRTBA and the parameterized rates is better at low electron densities, while at higher densities FT-pnRTBA returns lower EC rates than the systematics. The discrepancy can be attributed to the absence of the forbidden transitions in the present calculations and missing correlations of higher complexity. To verify the latter possibility, we explored the potential of the most recent approach which further extends the FT-pnRTBA with the GSC$_{\text{PVC}}$ correlations of higher complexity, that were found important for the description of GT$_+$ strength in neutron-rich nuclei. Although this approach is not yet generalized for finite temperatures, our estimate showed that these correlations should enhance the EC rates at high electron densities. 

In this way, the complex nuclear correlations beyond the one-loop approximation of the (FT)-QRPA type are found important and to be included in the calculations of EC rates in stellar environments.
In particular, the PVC correlations are necessary as they reproduce the GT strength distributions considerably better. However, in some cases the leading-approximation PVC correlations may be not sufficient for the accurate determination of the GT$_+$ excitations and EC rates in neutron-rich nuclei and more sophisticated correlations are needed. Further model developments as well as experimental studies of the GT$_+$ transitions in neutron-rich nuclei are in order to clarify the remaining issues.

\section{Acknowledgements}
We appreciate illuminating discussions with Alan Dzhioev, Francesca Gulminelli, Anthea Fantina, Michael Famiano and Remco Zegers. Financial support  by the US-NSF Career Grant PHY-1654379 is gratefully acknowledged.
%

\bibliography{Bibliography_Sep2020}
\end{document}